\documentclass[14pt,english]{article}
\usepackage[dvips]{graphicx}
\usepackage{babel}
\usepackage{inputenc}
\date{}
\title{\Large{THE HIGH ENERGY FRONTIER AND THE}\\
\Large{ COLLABORATION IN COLLIDER EXPERIMENTS}}
\author{\normalsize{G. GIACOMELLI} \\
\small{\emph{Physics Dept. of the University of Bologna and INFN,
Sezione di Bologna,}}\\
\small{\emph{Viale Berti-Pichat 6/2, I-40127 Bologna, Italy}}\\
\small{\emph{E-mail: giacomelli@bo.infn.it}}
}
\begin{document}
\maketitle

\begin{verse}
\begin{center}
{\bf{Invited paper at the Italy-Japan Int. Symposium,}}\\
{\bf{ Nagoya University, 15/10/2001}}
\end{center}
\end{verse}

\begin{center}
\begin{verse}
\hspace{0.5cm}In this note I shall review the Japan-Italy cooperative work
in high energy collider experiments, in particular OPAL at LEP,
CDF at Fermilab, ZEUS at HERA and ATLAS at the future LHC.
\end{verse}
\end{center}

\large{
\section{Introduction}
The Standard Model (SM) of particle physics has been checked to an
unprecedented level of accuracy by the precision measurements at LEP,
SLC, HERA and the Fermilab colliders. For the Higgs boson there were
some indications and limits at LEP: the search for the Higgs boson is
one of the main items to be studied at the Fermilab-2 $\overline{p}p$
collider and at the future LHC pp collider. The SM has theoretical
inconsistencies and too many free parameters, and it could be a low
energy approximation of a more complete theory. Thus many physicists
are looking for physics beyond the SM, for instance supersymmetry,
compositeness, etc. New particle searches is thus an other important
subject of research. Connected with these problems there is also a
strong interest in finding the deep structure of the proton at ever
smaller distances.

Experimentation at higher energies forced the concetration of
the experiments in larger national and international laboratories,
where high energy accelerators were available. From  the 1960's the
experiments started to grow in size and complexity and started to be
performed by collaborations of few groups from different Institutions:
experiments grew larger, with a tendency to go towards the use of all
purpose detectors. The experiments at LEP, SLC, Fermilab and HERA,
required another step, with tens of groups and hundreds of physicists
and engineers, with interconnections at the national and regional
levels. The next experiments at the LHC require hundreds of groups and
thousands of physicists, with interconnections in a sort of world
organization.

In this context many cooperative efforts were made by Japanese
and Italian groups.  Here I shall first recall a personal experience
in the United States at the end of the 1950's, and then I shall
discuss the cooperative efforts by Japanese and Italian groups in the
experiments OPAL at LEP, CDF at Fermilab, ZEUS at HERA and ATLAS at
the future LHC collider.

It may be worth recalling that these collider experiments use
4$\pi$ general purpose detectors, with many subdetectors, some of
which immersed in a magnetic field. They have several hundred thousand
eletronic channels and very many microprocessors and computers. The
detectors have a cylindrical symmetry, with a "barrel" and "endcap"
structure; some subdetectors are further structured in different
ways. Starting with the innermost detectors and proceeding outward one
finds: a microvertex detector, a central tracking detector with dE/dx
and time-of-flight capabilities; the momentum is measured by track
curvature in the magnetic field. Then follow the electromagnetic and
hadron calorimeters, and, after the iron of the magnetic field return
yoke, the muon detector.

The collider luminosity was measured using forward detectors
and precision "luminometers". 

\section{The "old times"}
In the 1950's the standard high energy experiment was small and was
performed by a small group of physicists and graduate students from a
single university.

At the University of Rochester, Rochester, N.Y., USA, a
special situation was created by the late Robert Marshak, who accepted
many foreing students, mostly from Japan and Italy. They attended
graduate classes together, participated to seminars and colloquia and
performed experiments together, in small groups. I like to recall some
of the Japanese and Italian colleagues: M. Koshiba, K. Miyake,
S. Okubo, T. Yamanouchi and others from Japan; from Italy: R. Cester,
G. Maltoni, T. Regge, C. Schaerf and others. This was probably one of
the first interactions between Japanese and Italian physicists. In
their diversity of culture they found many things in common, among
which the fact that both lost world war two!

The high energy accelerator of that time at Rochester was a
syncrocyclotron of 240 MeV, which produced pions, below
threshold. Some of the main fields of experimental research concerned
the measurement of the cross sections for pion-nucleon elastic and
charge exchange processes at lab. energies smaller than 100 MeV
\cite{barnes} and analyses of nuclear emulsions mainly exposed to
cosmic rays. There was also a very strong theory group. 

\section{OPAL at LEP}
OPAL (Omni Purpose Apparatus for Lep) was one of the four LEP
deterctors \cite{ahmet}; it used a conventional magnetic field. The
collaboration, which started in the early 1980's, consisted of about
300 physicists and graduate students from 34 different Institutions
from many different countries. About 20 Japanese physicists and
graduate students, mainly from the University of Tokyo, the
International Center for Elementary Particle Physics (ICEPP), and from
the University of Kobe, participated in OPAL. Approximately the same
number of physicists and students were from Italy, mainly from the
University of Bologna, the Bologna Section of INFN and CERN.

The Japanese collegues constructed and maintained the Barrel
electromagnetic calorimeter, made of 9440 lead glass counters, all
working after 11 years of operation! They also provided a large
computer (a mainframe), which was essential for off-line analyses at
the beginning of LEP operations.

It may be worth recalling that computer technology increased
very fast, about a factor of 1000 between LEP constuction in the
1980's and the end of LEP operations in November 2000. The tendency
was towards clusters of PCs. Off-line analyses involved off-line
reconstruction of events, Monte Carlo simulations and physics
analyses. During this period of about 15 years the storage capacity
went from megabytes to gigabytes, the computer interconnectivity
improved dramatically, and CERN made its most important invention:
WWW, the World Wide Web.

The Italians constructed and maintained the Iarocci limited
streamer tubes for the Hadron Calorimeter. The tubes were fabricated
at the "tubificio" in Frascati, and use was made of the so called
"magic oil", which was pressed in place with a sponge; this improved
the quality of the inner conducting surfaces of the tubes, which then
worked more reliably: the vast majority of the tubes was still working
well at the end of the experiment. Italy contributed also to the
Luminometer and to the microvertex detector.

During the first years of data taking, it became clear that
several improvements could be made. In particular: the available
forward detectors allowed luminosity measurements to few percent. In
order to fully exploit the LEP accelerator and the detectors,
precisions "Luminometers" were designed and built: they allowed
measurements with precisions of better than 0.1\%. To reach these
precisions it was also essential to compute the radiative corrections
of forward $e^{+}e^{-}$ Bhabha elastic scattering to ever increasing
precisions: there was a healthy competition between theorists and
experimentalists to reach the desired goal. The final measurements
were made with impressive precisions, much better than expected. The
same "silicon technology" was used for the new microvertex
detectors.

Both Italians and Japanese physicists were very active in
improving and analyzing the luminosity measurements, and in many types
of physics analyses.

\section{CDF at Fermilab}
The CDF (Collider Detector Facility) \cite{cdf} is one of the two
general purpose detectors operating at the Fermilab $\overline{p}p$
collider. CDF started taking data in the 1990's at a center of mass
energy of 1.8 TeV. In the second part of the 1990's there was a long
shutdown of the collider: several improvements were made resulting in
an energy increase from 1.8 to 2 TeV and to a considerable luminosity
increase. Also the CDF detecor had many improvements: it is now called
CDF2 and it started the so called run2. The CDF collaboration consists
of about 480 physicists and graduate students from $\sim$ 50
Institutions from all over the world. The Japanese physicists are
about 50 from 5 Institutions (Hiroshima, KEK, Osaka, Tsukuba, Waseda),
while the Italians are about 80 from 6 Institutions (Bologna,
Frascati, Padova, Pisa, Roma 1, Trieste).

Also in this experiment the Japanese and Italian contributions
were and are very important: Japanese and Italians built different
crucial subdetectors; in particular they contributed to the equipment
made for the precision measurements and to the microvertex detectors;
the Italian groups pioneered the silicon microstrip detectors and
constantly improved them.

CDF contributed to the precision electroweak measurements,
discovered the top quark, studied in detail quark-quark collisions
yielding large $p_t$ events, studied minimum bias events, b-physics
and performed many new particle searches.

The EW precision measurements from LEP, SLC and Fermilab
yielded the following values \cite{lep}:  

\begin{figure}[h]
{\centering\resizebox*{!}{8.3cm}{\includegraphics{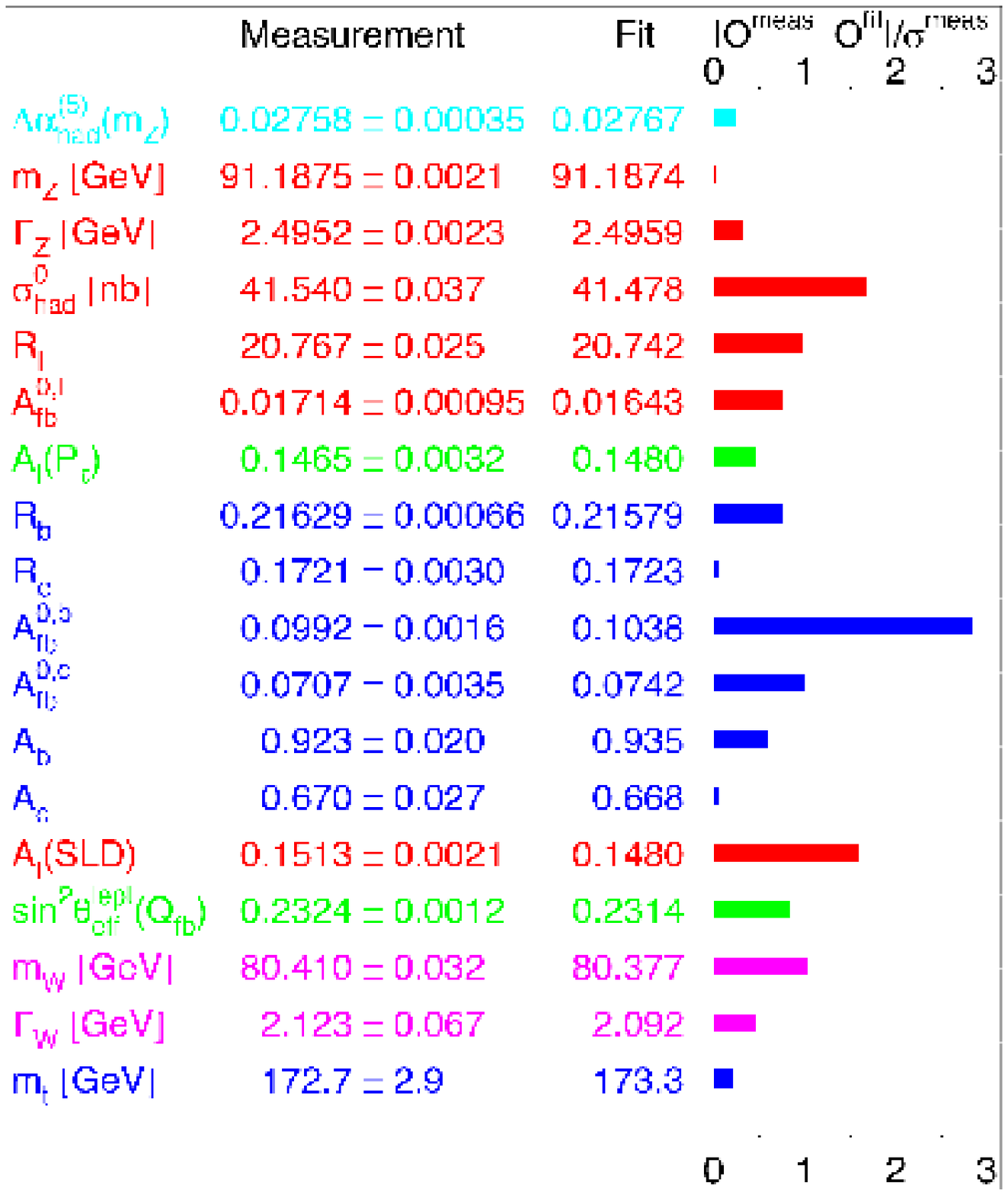}}\par}
\end{figure}

\section{ZEUS at HERA}
The HERA electron-proton collider is an asymmetric collider, with 30
GeV positrons (or electrons) in one ring and 820 GeV protons in the
other ring; the c.m. energy is about 300 GeV. There was a recent
upgrade which resulted in an increase of the HERA luminosity.

The ZEUS detector is one of the two general purpose detectors
at HERA \cite{zeus}. The ZEUS collaboration is formed by about 460
physicists, engineers and students from 52 Institutions from many
countries.

The Japanese participation in HERA concerns about 20
physicists and students from 5 Institutions; there are about 55
italian physicists and students from 7 Institutions. The Japanese
concentrated in the electromagnetic calorimeter and in the problem of
hadron-electron separation. The italian contribution concerned the
muon detection system, the leading forward (proton) spectrometer and
the silicon microvertex detectors. Italy also contributed considerably
to the construction of HERA.

There was a recent upgrade of the ZEUS detector (in particular
of the silicon microvertex).

ZEUS results concern the study of the proton structure
function, $F_2$, from data on deep inelastic scattering at very low
values of the variable x and at very large values of $Q^2$, the study
of the photon structure, of heavy flavour production, of leading
proton physics, of hadronic final states, of many searches for new
particles and new phenomena, etc. 

\section{ATLAS at LHC}
The future LHC collider at CERN will accelerate protons to 7 TeV, thus
the c.m. energy will be 14 TeV. It will have a very high luminosity,
about $10^{34}~cm^{-2}s^{-1}$. This will correspond to about 20
collisions per bunch crossing, which will have to be unscrambled by
fast timing and fast triggers. 

The ATLAS (A Toroidal LHC Apparatus) detector \cite{atlas} is
one of the general purpose detectors at LHC; more than 1500 physicists
and graduate students are building equipment and perform Monte Carlo
simulations of complicated events [each of which requires a very large
computing power]. The physicists are from 150 Institutions from 34
Nations [plus CERN + Dubna] from the 5 continents. It is thus a world
collaboration, and it requires a complex organization.

There are about 50 physicists and students from 12 Japanese
Intitutions in ATLAS; from Italy there are about 100 physicists and
students from 12 Institutions. The Japanese colleagues are involved in
the superconducting solenoid, in the end-cap chambers, in electronics,
triggers and other items. The Italians are involved in the B0 and
toroid magnetic systems, in the RPC detectors for the barrel, in
tracking detectors, and in calorimeters.

A major common effort of the LHC collaborations concerns the
GRID computing project for off line analyses and Monte Carlo
production \cite{capiluppi}; it also aims to improve the efficiency,
speed and connectivity of all computers systems.

The expected physics programs at LHC will first concentrate on
the search for the Standard Model Higgs boson; other studies will
concern the physics items already discussed in Sections 3-5, in
particular on new particle searches of larger masses, especially
supersymmetric particles, using the anticipated very high
luminosities. Other studies will concern heavy quark physics, large
and small $p_t$ physics, also heavy ion collisions, etc.

\section{Conclusions}
It seems that most physicists who have been involved in one of the
large collider experiments consider their experience to have been
exciting. This was more so when the experiments were obtaining their
most interesting physics results, like the 3 neutrino families, the
precision determination of the electroweak parameters and of the
strong interaction (QCD) parameters [3], the running of the strong and
of the electromagnetic coupling constants, precise measurements of the
lifetimes of short lived particles, the existence of the triple boson
vertex, the discovery of the top quark, possible indications and new
limits on new particles \cite{barate} and new phenomena.

The contributions of Japanese and Italian collegues were
always very important in the planning, construction and physics
analyses.

One should not neglect the very large number of Diploma,
Laurea and PhD theses with data from high energy collider experiments,
and the strong impact on the public understanding of science.\\ 

I thank many colleagues from the OPAL, CDF, ZEUS and ATLAS
collaborations for their cooperation, in particular G. Bellettini,
H. Bilokon, P. Capiluppi, S. Zucchelli and many others. I thank
Drs. A. Casoni and G. Grandi for typing the manuscript.

}
\end{document}